\documentclass[preprint,prl,aps,amsfonts,amssymb,amsmath,showpacs]{revtex4}
\usepackage{graphicx}

\begin{document}
\bibliographystyle{prsty}
\title{Implications of protective measurements on de Broglie-Bohm trajectories}

\author{Aur\'{e}lien Drezet }
\affiliation{Institut N\'eel, CNRS et Universit\'e Joseph Fourier,
BP 166, F-38042 Grenoble Cedex 9, France}

\date{\today}

\begin{abstract}

\end{abstract}

\maketitle

\section{Motivations}
Protective measurements which were defined by Aharonov and Vaidman
in 1993 \cite{Aharonov1993a} played an important role in the
discussion about the interpretation of quantum mechanics. In 1999,
following an early work by Englert et al.\cite{Englert1992} Aharonov
et al.~\cite{Aharonov1999} wrote an article in which they showed
that protective measurements can be used to demonstrate the
`surrealism' of Bohmian mechanics. Bohmian mechanics also known as
pilot-wave interpretation is certainly the best-known hidden
variable interpretation of quantum mechanics. It played a
fundamental role in the discovery by Bell of his famous non locality
theorem. Therefore, any attacks against pilot-wave interpretation is
particularly interesting and instructive to learn something new
about the mysterious quantum universe. It is the aim of this chapter
to review the debate surrounding protective measurement and
pilot-wave (see also \cite{Drezet2005} and \cite{Gao}) and to show
if it is possible to reconciliate the different interpretations of
the results given in \cite{Aharonov1999}.
\section{An historical review of pilot-wave interpretation}
We first remind to the reader some basics about de Broglie-Bohm
`pilot-wave' ontology and in particular about its curious history.
De Broglie proposed his approach to quantum mechanics in the period
1925-1927, i.e., at the early beginning of modern quantum physics as
we know it. De Broglie based his interpretation mainly on
relativistic considerations and discovered along this path what is
nowadays known as the `Klein-Gordon' equation:
\begin{equation}
\Box\Psi(\mathbf{x},t)=-\frac{m_0^2}{\hbar^2}\Psi(\mathbf{x},t),\label{1}
\end{equation}
What is however puzzling is that the first calculations he did on
this subject in 1925 \cite{broglie1925a} were realized before the
discovery by Schr\"{o}dinger of his famous equation. In some way, we
can therefore say that it is quantum wave mechanics which was a
development of pilot-wave theory and not
the opposite~\cite{Valentinibook}.\\
More precisely, the starting idea of de Broglie \cite{broglie1925b}
was that each single quantum object is actually some highly
localized singularity of a specific wave field $\Psi(\mathbf{x},t)$
which should ultimately be solution of a yet unknown non-linear wave
equation. Following Einstein, which had already proposed similar
ideas in 1909~\cite{Einstein1909} for photons (the so called
`Nadelstrahlung' concept), de Broglie started a research program
baptized `double solution'~\cite{broglie1927} in which each quantum
is some `bunched' oscillating region of the field propagating as a
whole like a particle (i.e. in modern words: a soliton) and inducing
a much weaker wave field in its surrounding. This weaker field was
supposed to be in `harmony of phases' with the singular field so
that both were locked to each other. Following this program the
weaker field should obey, far away from the core, a linear equation,
e.g., Eq.~\ref{1}, and subsequently should act as a guiding or pilot
wave for the singular part, i.e., determining his complete dynamics.
This was of course a very ambitious project and not surprisingly de
Broglie never succeeded to complete his theory~\cite{broglie1956}.
Still, during his early quest in 1927 he found a `minimalist
solution' which is the foundation of what we call nowadays the de
Broglie-Bohm interpretation of quantum mechanics. The theory was
introduced at the end of a long article about his double solution
program~\cite{broglie1927} and was subsequently presented during the
5$^{th}$ Solvay congress which took place in Brussels~\cite{Solvay}
(p. 105-132). In pilot-wave mechanics, the wave is everywhere
reduced to its linear contribution, e.g., a solution of
Schr\"{o}dinger equation in the non relativistic regime. The
particle behave like a point-like object whose motion is completely
determined by the linear wave. De Broglie was able to define the
equation of motion of the moving point like particles (for the
single and many electron cases)
and showed how to solve the dynamic for some specific problems.\\
\indent Consider for example a single electron described by
Schr\"{o}dinger's equation:
\begin{equation}
i\hbar\frac{\partial}{\partial
t}\Psi(\mathbf{x},t)=\frac{-\hbar^2}{2m_0}\triangle\Psi(\mathbf{x},t)+V(\mathbf{x},t).\label{2}
\end{equation}
If we know a solution of this equation written in polar form as
$\Psi(\mathbf{x},t)=a(\mathbf{x},t)e^{iS(\mathbf{x},t)/\hbar}$ we
can define a density of probability
$\rho(\mathbf{x},t)=\Psi(\mathbf{x},t)\Psi(\mathbf{x},t)^\ast$
,i.e., $\rho(\mathbf{x},t)=a(\mathbf{x},t)^2$ and a probability
current $\mathbf{J}(\mathbf{x},t)$ such as
\begin{eqnarray}
\mathbf{J}(\mathbf{x},t)=\hbar\frac{\Psi(\mathbf{x},t)^\ast\nabla\Psi(\mathbf{x},t)-\Psi(\mathbf{x},t)\boldsymbol{\nabla}\Psi(\mathbf{x},t)^\ast}{2i
m_0}=a(\mathbf{x},t)^2\frac{\boldsymbol{\nabla}S(\mathbf{x},t)}{m_0}.
\end{eqnarray}
Using these equations de Broglie defined the velocity of the
particle as
\begin{eqnarray}\mathbf{v}(t)=\frac{d}{dt}\mathbf{x}(t)=\frac{\mathbf{J}(\mathbf{x},t)}{\rho(\mathbf{x},t)}=\frac{\boldsymbol{\nabla}S(\mathbf{x},t)}{m_0},\label{4}
\end{eqnarray} showing that in analogy with classical dynamics
$S(\mathbf{x}(t),t)$ plays the role of an action (see also
Madelung~\cite{Madelung1927}). This analogy is even enforced when we
insert $a$ and $S$ in Eq.~\ref{2} to obtain
\begin{eqnarray}
-\frac{\partial}{\partial
t}S(\mathbf{x}(t),t)=\frac{(\boldsymbol{\nabla}S(\mathbf{x},t))^2}{2m_0}+V(\mathbf{x}(t),t)-\frac{\hbar^2\triangle
a(\mathbf{x}(t),t)}{2m_0a(\mathbf{x}(t),t)}.\nonumber\\
\label{hamilton}
\end{eqnarray}
We recognize the well-known Hamilton-Jacobi equation  which in
classical dynamics determines the motion of the particle in an
external potential $V$. However, there is here an additional term
$Q(\mathbf{x},t)=-\hbar^2\triangle
a(\mathbf{x},t)/(2m_0a(\mathbf{x},t))$ called the quantum potential
by de Broglie. This potential is determined by the wave amplitude in
agreement with the pilot-wave idea. Importantly $Q$ is unchanged if
the wave function is multiplied by a constant so that actually it is
the form of the wave more that its amplitude which has a
signification in this theory.  Also, for a many body system the
potential $Q (x_1,x_2,...x_N,t)$ depends in general in a nonlocal
way of the  $N$ particle coordinates.  This can lead to some
specific features such as non local entanglement discussed in the
context of the EPR paradox~\cite{EPR} or the Bell
inequality~\cite{Bell}. In particular, the fact that pilot-wave
theory agrees with Bell's theorem implies some kind of mysterious
action at a distance between the particles. We point out that de
Broglie contrary to Bohm was very reluctant to introduce non
locality in his ontological theory and that he expected to remove
this feature with
his double solution program.\\
\indent Importantly, the Hamilton-Jacobi analogy suggests that
pilot-wave theory can equivalently be written in Newton's form. The
second law for de Broglie's dynamics is indeed easily written as
$m_0\frac{d^2}{dt^2}\mathbf{x}(t)=-\boldsymbol{\nabla}[V(\mathbf{x}(t),t)+Q(\mathbf{x}(t),t)]$
in full analogy with classical dynamics for a point-like particle.
However, while this dynamical law contains a second order time
derivative it is important to observe that for practical purposes if
$\Psi(\mathbf{x},t)$ is known then the first order Eq.~\ref{4} is
sufficient to completely describe the trajectories. This is indeed
done through integration of the flow equations:
\begin{equation}
\frac{dx}{\frac{\partial S(\mathbf{x},t)}{\partial
x}}=\frac{dy}{\frac{\partial S(\mathbf{x},t)}{\partial
y}}=\frac{dz}{\frac{\partial S(\mathbf{x},t)}{\partial
z}}=\frac{dt}{m_0}.\label{6}
\end{equation} for a given initial condition
$\mathbf{x}(t_0)=\mathbf{x}_0$. This point is important  because
John Bell~\cite{Bell} used pilot-wave theory mainly through the
definition given by Eq.~\ref{4} while other authors  like
Vigier~\cite{Vigier1954} and Bohm~\cite{Bohm1952a} insisted on the
need to use the quantum potential for a complete physical
description of the particle motion. This seems to indicate that
the theory lacks a univocal axiomatic for his foundation.\\
\indent At Solvay conference W.~Pauli was probably the most reactive
concerning criticisms but even potential followers like Einstein or
the more `classical' Lorentz were not showing a too strong
enthusiasm for de Broglie pilot-wave approach. Remarkably, due to
internal mathematical difficulties of his `double solution' program
de Broglie only presented his pilot-wave version in Brussels. This
was certainly an honest choice but physically far less profound and
less impressive for this demanding audience. In particular, one of
Pauli's objection concerned the arbitrariness of the dynamics law
obtained by de Broglie.  Indeed, Pauli observed~\cite{Solvay} (see
p. 134-135) that the dynamics proposed by de Broglie has no precise
foundation since the conservation current is not univocal, i.e. one
can add a divergence free vector to $\mathbf{J}$ without changing
the conservation law, and Schr\"{o}dinger asked further why we
should not use instead of Eq.~\ref{4} a different
definition~\cite{Solvay} (p.~135), e.g., the energy-momentum tensor
$T^{\mu\nu} (\mathbf{x},t)$ in order to define a trajectory. This
was indeed proposed by de Broglie himself for
photons~\cite{broglie1926} (see however \cite{Holland} for a modern
perspective concerning this problem and the
difficulties about a covariant generalization of pilot-wave).\\
\indent We also mention a related critical comment concerning
foundation made in 1952 by Pauli~\cite{Pauli} and in 1955 by
Heisenberg~\cite{Heisenberg}. Both physicists indeed complained by
observing that for de Broglie and Bohm the particle position plays a
fundamental role that breaks the accepted a symmetry between
position and momentum (symmetry which is at the heart of quantum
formalism). This was unacceptable for Heisenberg, and Pauli, for
whom position $q$ and momentum $p$ should be introduced at an
equal footing.\\
\indent Of course, all these observations by Pauli, Schr\"{o}dinger
and Heisenberg are not decisive remarks against pilot-wave
interpretation since the plausibility or un-plausibility of the
dynamics doesn't constitute by itself a proof or disproof of the
theory: only experiments should have the last word. Nevertheless,
altogether these problems let a strong feeling of discomfort to the
audience of the Solvay conference and to the first generations of
quantum theorists. This discomfort never really disappeared until
nowadays.
\section{The measurement theory and the adiabatic theorem}
 \subsection{Einstein's reaction}
Beyond these interesting problems about axiomatics and foundations
the most critical part of the theory concerns of course his
agreement with experimental facts and the realism of the predictions
given by the pilot-wave approach. Indeed, if Schr\"{o}dinger's
equation completely determines the particle motion through
Eq.~\ref{4} then we expect that both the usual `Copenhagen' approach
and the one of de Broglie should be experimentally equivalent. This
was indeed later confirmed after the more detailed studies of
measurement processes by David Bohm in 1952~\cite{Bohm1952b}. Still,
in 1927 de Broglie \cite{broglie1927} already showed that the
four-vector current $J^{\mu}$, which naturally arises from wave
equation \ref{1} and formally leads to the conservation law
$\partial_\mu J^{\mu}=0$ through the Noether theorem, can be used to
justify the statistical interpretation of quantum mechanics, i.e.,
the so called `Born's probability rule'. Indeed, if for simplicity
we limit ourself to the non relativistic regime then, the evolution
equation Eq.~\ref{4} and the current conservation rule
$\partial_t\rho+\boldsymbol{\nabla}\cdot\mathbf{J}=0$, imply the
following: if at a given time the probability distribution of
particle in space is given by $a^2$ then this will also be true at
any time. If we write $a^2(\mathbf{x}(t_0),t_0)$ the density of
probability at $\mathbf{x}_0=\mathbf{x}(t_0)$ and time $t_0$ we can
obtain by direct integration the density of probability at time $t$
for the point $\mathbf{x}(t)$ located along the de Broglie
trajectory (see Eqs.~\ref{4},\ref{6}). We get
\begin{equation}
a^2(\mathbf{x}(t),t)=a^2(\mathbf{x}(t_0),t_0)\cdot
e^{-\int_{t_0}^{t}dt'\frac{\triangle'S(\mathbf{x}(t'),t')}{m_0}},
\end{equation} where $\triangle'=\partial^2 /\partial
\mathbf{x}(t')^2$. This is the same reasoning which is used in
classical statistical mechanics, e.g., Liouville, Gibbs, to justify
probability laws. In particular, if at given time the wave functions
of particles can be approximated by uncorrelated plane waves then
and `apriori' symmetry implies the homogeneity  of probability
distribution in space (this can be seen as an initial chaotic
condition \`{a} la Bolstzmann, i.e., a `Stosszahlansatz'). The
subsequent interaction processes between the different particles
will certainly create correlations between them but then the
deterministic evolution (e.g., Eq.~\ref{4} and its generalization
for the many-body problem) will maintain the probability
interpretation for any other time $t$ as we already said. This idea
was further developed by Bohm~\cite{Bohm1953},
Vigier~\cite{Vigier1954} and Nelson~\cite{Nelson1966} in the 50-60's
and more recently by
Valentini~\cite{Valentini1991a};\cite{Valentini1991b}, D\"{u}rr,
Goldstein
and Zanghi~\cite{Durr1992} with different strategies.\\
\indent This is certainly impressive, or at least promising, but the
theory possesses some other `repellant' features which were studied
in the recent years and are the subject of the present chapter. One
of them already mentioned by Ehrenfest in~\cite{Solvay} p. 136
concerns the fact that in the ground state of an Hydrogen atom (i.e.
a s state) the wave function is (up to the $e^{-iEt/\hbar}$
contribution) real. It implies
$\mathbf{v}=\boldsymbol{\nabla}S/m_0=0$ i.e, the fact that the
electron is at rest in the s-atom. From the point of view of the de
Broglie theory there is nevertheless no contradiction since the
constant energy $E$ is given by $E=-\partial_t S=
V(\mathbf{x})+Q(\mathbf{x})$ and the variation of $Q$ with
$\mathbf{x}$ exactly compensates the variation of $V$. The force
$\mathbf{F}=-\boldsymbol{\nabla}[V+Q]$ therefore vanishes and the
electron is not accelerated. Still, this feature looked not
realistic and played again against de Broglie.  Not surprisingly,
after this period 1927-1928 de Broglie abandoned his theory and went
back to it only after 1952 and the rediscovery by Bohm of
pilot-waves. We point out that the
`$\mathbf{v}=\boldsymbol{\nabla}S/m_0=0$' objection played also a
role in the `cold' reception of this theory by Pauli~\cite{Pauli}
and Einstein and this even after 1952 (see also
Rosen~\cite{Rosen1945} who re-discovered, after de Broglie but
before Bohm, the pilot-wave concept and repudiated it for the same
reasons as Einstein). In a paper written for Max Born retirement
from the University of Edinburgh~\cite{Born} Einstein discussed the
example of a particle in a infinite 1D potential well which admits
wave functions\begin{equation}\Psi(x,t)=\sqrt{\frac{2}{L}}\sin{(n\pi
x/L)}e^{-iE_nt/\hbar}\label{9}\end{equation} associated with the
energy $E_n=(\hbar n\pi/L)^2/(2m_0)$ for $n=0,1,2,$ etc. Clearly,
here again the velocity of the particle cancels.  For Einstein this
seemed to contradict the fact that for large $n$  an ontological
theory like pilot-wave should `intuitively' recover classical
mechanics. However, in classical mechanics we have $Q=0$ and
$E=p^2/(2m_0)$ with $p=m_0v$. This apparently fits well with
Schr\"{o}dinger equation if we write $p=\hbar n\pi/L$.
Unfortunately, pilot-wave of de Broglie and Bohm implies
$p=m_0v=\nabla S=0$ and $Q=(\hbar n\pi/L)^2/(2m_0)=E$. Most
remarkably, this occurs independently of how large the quantum
number $n$ is and is therefore in complete contradiction with what
we intuitively expect in the classical regime. Commenting further on
Bohm's attempt to reintroduce pilot-wave theory Einstein once wrote
to Born `That way seems too cheap to me'. Still, we point out that
neither de Broglie  nor Bohm agreed with Einstein's conclusion. For
example, in his book written with B.~Hiley~\cite{Hiley} Bohm replied
that, independently of the details of pilot-wave theory, any model
attempting to preserve the particle localization in the infinite
potential well would ultimately contradicts classical physics. This
should be the case since at fixed energy there are necessarily some
nodes where the wave function cancels and are therefore prohibited
to the particle localization, i.~e., corresponding to regions where
the probability is zero. In the 1D case the potential well is thus
obviously separated into small spatial cells of size $\lambda/2=L/n$
where the particle is confined and cannot escape because it cannot
cross or even reach the nodes. Therefore, in this context the
expectation of Einstein appears illusory. Still, the example of
Einstein or the one of the s atom constitute perfect illustrations
of the `surrealistic nature of de Broglie-Bohm trajectory'. This
qualifier was given by Englert \emph{et al.} after a very detailed
paper~\cite{Englert1992} which discussed pilot-wave interpretation
in the context of measurement theory.
\subsection{Von Neumann's strong measurements}
The most important contribution of David Bohm to pilot-wave theory
concerns his interpretation of quantum measurements. In 1952 in a
series of two well known papers \cite{Bohm1952a};\cite{Bohm1952b} he
discussed the canonical von Neumann projective measurements in the
context of pilot-wave theory.  He showed that there is nothing of
contradictory or impossible in attributing at the same time a
position and a momentum to a particle as soon as we accept that the
so-called momentum measured is not in general its actual momentum.
This should be already clear from the definition
$\mathbf{p}(t)=m_0\mathbf{v}(t)=\boldsymbol{\nabla}S$ which holds at
any location $\mathbf{x}$ visited by the particle. The plane-wave
eigenstates $|\mathbf{p}\rangle$ of the operator
$\hat{\mathbf{p}}=-i\hbar\boldsymbol{\nabla}$ are completely
delocalized and according  to Heisenberg principle this prohibits a
clean localization of the particle. The agreement between the
definition of de Broglie-Bohm on the one side and of Heisenberg on
the other side  is reestablished if we realize that in order to
measure the momentum associated with the operator $\hat{\mathbf{p}}$
one must disturb the initial wave function and separate the
different plane wave contributions.\\ Consider once again the
example of the infinite potential well. Bohm observes that the wave
function given by Eq.~\ref{9} can be formally expanded into plane
waves and the Fourier amplitudes $\tilde{\psi}(p)$ correspond to two
well localized wave packets peaked near the momentum values $p=\pm
\hbar n/L$ (in the classical  limit $n\rightarrow +\infty$ we have
$|\tilde{\psi}(p)|^2\simeq [\delta(p-\hbar n/L) +\delta(p+\hbar
n/L)]/2$). Bohm then supposed that the walls or the well confining
the particle instantaneously disappears without disturbing in any
appreciable fashion the wave function. The two wave packets
subsequently propagate freely and ultimately separate from each
other. This makes the packets spatially distinguishable and allows
for a measurement of the particle momentum $p\simeq\pm \hbar n/L$.\\
\indent This example is of course a `gedanken' experiment and
subsequent studies made by Bohm and followers focussed on the
procedure of entanglement between a pointer or meter and the
analyzed quantum system.\\
This was first done in the context of von Neumann measurement whose
method was well discussed by Bohm himself in his `orthodox' 1951
text book, e.g., for the Stern-Gerlach experiment
analysis~\cite{Bohm1951}. The main idea can be easily illustrated by
considering the total Hamiltonian
\begin{eqnarray} \hat{H}(t)=\hat{H}_S+\hat{H}_M -\hbar
g(t)\epsilon\hat{A}_S\hat{X}_M,
\end{eqnarray} describing the interaction between a system S
and a meter M.  The operator $\hat{A}_S$ acts only on S and
corresponds to the variable we wish to measure. $\hat{X}_M$ is the
operator describing the meter. It represents here its position (e.g.
the atom center of mass in the Stern-Gerlach experiment). The
coupling is also characterized  by a  constant $\epsilon$ introduced
for the sake of the equation homogeneity and a time dependent
function $g(t)$ characterizing  the  fast evolution of the
measurement protocol. Here, we impose to simplify $g(t)=\delta(t)$,
i.e., an instantaneous measurement. Before the interaction occurs at
$t=0$ we start, i.e., for $t<0$, with two decoupled and unentangled
subsystems S and M described by the quantum state
$|\Psi_{in}(t)\rangle=|S(t)\rangle\otimes|M(t)\rangle$. After the
interaction occurred , i.e. for $t>0$, we obtain the final state
$|\Psi_{f}(t)\rangle=\hat{U}_S(t,t=0)\hat{U}_M(t,t=0)|\Psi_{f}(0)\rangle$
where $\hat{U}_S(t,t=0)$ and $\hat{U}_M(t,t=0)$ are the evolution
operators of the freely moving subsystems S and M acting on
\begin{eqnarray}
|\Psi_{f}(0)\rangle=e^{i\epsilon\hat{A}_S(0)\hat{X}_M(0)}|S(0)\rangle\otimes|M(0)\rangle
\nonumber\\=\sum_a\int dp S(a)M(p) e^{i\epsilon
a\hat{X}_M}|a\rangle\otimes|p\rangle\end{eqnarray} or equivalently
on
\begin{eqnarray}|\Psi_{f}(0)\rangle =\sum_a\int dp
S(a)M(p)|a\rangle\otimes|p+\hbar\epsilon a\rangle
\nonumber\\=\sum_a\int dp S(a)M(p -\hbar\epsilon
a)|a\rangle\otimes|p\rangle.
\end{eqnarray}Here we used the expansion of
the initial wave packets in  the vector basis $|a\rangle$ and
$|p\rangle$ respectively and applied well-known properties of the
translation operators $\hat{T}(\hbar a)=e^{i\epsilon a\hat{X}_M}$.
In particular if we take $M(p)=e^{-\Delta p^2}$ we obtain after the
measurement a series of shifted gaussians $M'(p)=e^{-\Delta
(p-\hbar\epsilon a)^2}$ entangled with each state $|a\rangle$. If
the shift of each gaussian is larger than their typical width
$\delta p=1/(2\Delta)$ (and if we can neglect the free space
spreading of the pointer wave packets) it will be possible to
correlate the distribution $|S(a)|^{2}$ of S with the distribution
of gaussian centers in the momentum space of M. This is the basis of
the von Neumann measurement protocol which was translated into the
ontological language of pilot-wave theory by Bohm in 1952. For Bohm
indeed, the entanglement directly affected the particles
trajectories of the two subsystems S and M but if we observe the
meter at a location near $p\simeq \hbar\epsilon a$  it does nott
however implies that S is actually in the state $a$. This apparently
paradoxical result comes from the fact that in pilot-wave theory the
position of particles plays  a more fundamental role that in the
usual interpretation. Therefore, one should be authorized to speak
about measurement only if we can correlate the studied variables $a$
with the actual position of the system S. Interestingly, both
interpretations by von Neumann and Bohm of the previous protocol
will however eventually agree if the different wave packets of the
subsystem S: $\psi_a(x_S)$ in the base $a$ are not spatially
overlapping. In a more general way, if the entanglement between the
system S and meter M produces after the interaction a sum of
entangled states $\sum_{i}c_i\psi_i(x_S)\phi_i(x_M)$, where the
different wave functions for both particles are non overlapping, we
will then unambiguously be able to correlate the positions of S and
M with the states labeled by $i$. For most experiments this is
however not the case and the so-called quantum measurement cannot be
considered as such in the context of pilot-wave theory. It is
therefore amazing to observe that the famous dictum of Wheeler `No
elementary phenomenon is a phenomenon until it is an observed
phenomenon' which was given in the context of Bohr's interpretation
finds also his plain significance in the interpretation of de
Broglie and Bohm. Paraphrasing Wheeler, we could then state that `No
measurement is a measurement until it is a position measurement'. It
is also worth mentioning that in the same texts quoted previously,
both Pauli~\cite{Pauli} and Heisenberg~\cite{Heisenberg} criticized
this strange feature of pilot-wave approach. Heisenberg, in
particular, pertinently commented that in the deterministic approach
of Bohm momentum and position are in general hidden and correspond
therefore to metaphysical superstructures without any physical
implication.
\subsection{Protective measurements}
The previous discussion done in the context of orthodox von Neumann
strong projective measurements was extended in 1999 to the so-called
weak protective measurement domain by Aharonov, Englert, and Scully
in a fascinating paper~\cite{Aharonov1999}. The authors showed that
in the considered regime the interpretation by pilot-wave of the
results implied some even more drastic surrealism as in the strong
coupling regime. To understand their motivation it is important to
go back once again to the origin of pilot-wave mechanics  and to
observe that if the wave guides the particle during its motion then
in some situations empty waves without particle should exist.   For
example, in the double slit experiment  the particle travels through
one hole but something should go through the second hole in order to
disturb the motion on the other side and induce interference. Of
course, one can always involve the quantum potential as an
explanation but then one should explain why this potential exists
and the problem is therefore not removed. Many authors thinking
about this problem claimed that the guiding wave should carry some
energy and the particle should get less and less energy while
crossing an interferometer with more and more gates and
doors~\cite{broglie1956}. But obviously, this is not what is
predicted neither by quantum mechanics nor pilot-wave theory.
Another point, was that if empty wave reacts on the particle during
the double-slit interference experiment why should it not also acts
on some other systems~\cite{Selleri1990}. Could we detect an empty
wave? While working on this problem it was realized by
L.~Hardy~\cite{Hardy1992a};\cite{Hardy1992b};\cite{Hardy1992c} that
empty waves can sometimes have a physical effect on a second
entangled (measuring) system (his idea was actually an adaptation of
Elitzur and Vaidman `interaction free-measurement'
protocol~\cite{Elitzur1993}) and he found during his research a very
fascinating Bell's theorem without inequality involving strange non
local features and questioning the possibility to build up a Lorentz
invariant hidden variable model. The result of Hardy is intriguing
and also disappointing since, again, it is an indirect effect on
hidden variables which is observed. The empty wave affects the
dynamics of the second system but one must watch correlations
between events to see it (otherwise one could send faster-than-light
signals with this nonlocal protocol). For those already not
convinced by pilot-wave approach this definitely could not help. In
a different but related context J. Bell in 1980 \cite{Bell} (p.
111-116) studied the exotic behavior of Bohmian particles diffracted
by a screen and interacting with a complex detecting `which-path'
device. It was shown that the path followed by the particle is
sometime completely surrealistic and can even reach the wrong
detector (this is connected to the fact that Bohmian trajectory
cannot cross in the configuration space). However, this cannot
affect the interpretation since this is again hidden and impossible
to test experimentally. In a subsequent paper by Englert et
al.~\cite{Englert1992}, already mentioned (see
also~\cite{Dewdney1993} and \cite{Vaidman2012}) it was shown that
the problem is deeper than Bell thought at first, and that this
surrealism exists even with simple particles interacting with Stern
and Gerlach devices~\cite{Scully1998}. Therefore, to quote the
authors :`the reality attributed to Bohm trajectories is not
physical it is metaphysical'~\cite{Englert1992}. Lev
Vaidman~\cite{Vaidman2005} wrote once a very pedagogical paper
provocatively untitled: `The reality in Bohmian quantum mechanics or
can you kill with an empty wave bullet'. In his paper, Vaidman
explained with very symptomatic and illustrative examples (such as
slow bubble traces developing after the passage of the particle even
when the particle is not here but elsewhere) that if one is living
in pilot-wave world then entanglement with meters and environment
will break all your
convictions about causality and localization (i.e. in agreement with Hardy's conclusions).\\
\indent His paper reviewing the argument presented in
\cite{Aharonov1999} showed also that if the empty waves are involved
in all these processes then one can actually measure an empty wave
function without the particle being. This relies on protective
measurements of position with allow a measure of the wave function
density $|\psi(x)|^2$ of the particle at $x$ even if pilot-wave
trajectory never crosses the interaction region centered on $x$. The
concept of protective measurement is a beautiful idea which was
introduced by Y.~Aharonov and
L.~Vaidman~\cite{Aharonov1993a};\cite{Aharonov1993b} (see also
\cite{Aharonov1995} and \cite{Vaidman2005}). The principle relies on
the possibility to couple adiabatically the measuring device M with
the subsystem S in such a way to induce no significant change in the
$|S(0)\rangle$ initial state while disturbing the meter state
$|M(0)\rangle$ in an observable fashion. In such an approach, the
system S is therefore protected and it is easily shown that one can
use this kind of protocol to record an information on some local
observable such as $|\psi(x)|^2$ or $\mathbf{J}(x)$. The specific
example considered in \cite{Aharonov1999} is based, once more, on
the infinite potential well but now with a very local interaction
with a meter at one point (i.e. $0<x=x_0<L$) of the cavity.  The
total Hamiltonian is
\begin{equation}
\hat{H}(t)=\frac{-\hbar^2}{2m}\frac{\partial^2}{\partial x^2} +
\frac{-\hbar^2}{2M}\frac{\partial^2}{\partial X^2}-\hbar\epsilon
g(t)\delta(x-x_0)X
\end{equation} where  $x$ is the coordinate of the particle of mass $m$ in the box while $X$ is the coordinate of the
meter with mass $M>>m$. The coupling is monitored  by the external
parameter $g(t)$ such as $\int_{-\infty}^{+\infty} dt g(t)=1$. If
$g(t)$ changes very fast one goes back to the von Neumann regime but
here $g(t)$ changes very slowly, i.e., adiabatically, and it
vanishes outside the interval $[-T/2,+T/2]$ where it has the typical
value $g(t)\simeq 1/T$. The most characteristic feature of this
interaction is of course the presence of the Dirac function which
implies a short-range coupling existing only in the vicinity of
$x=x_0$. In order to solve the dynamical equation we apply here the
adiabatic approximation method~\cite{Bohm1951} and we first search
for eigenstates of the equation
$\hat{H}(t)\Psi(x,X,t)=E(t)\Psi(x,X,t)$. Inserting the `ansatz'
$\Psi(x,X,t)=\phi_s(x,X,t)e^{iPX/\hbar}/\sqrt{(2\pi\hbar)}$ we get
the new equation:
\begin{equation}
[E(t)-\frac{P^2}{2M}]\phi_s(x,X,t)=
\frac{-\hbar^2}{2m}\frac{\partial^2}{\partial
x^2}\phi_s(x,X,t)-\hbar\epsilon
g(t)\delta(x-x_0)X\phi_s(x,X,t).\label{14}
\end{equation} This is actually a 1D Green function problem with $t$
and $X$ as parameters. and Aharonov et al. solved it
analytically~\cite{Aharonov1999}. Still,  since we suppose the
coupling to be weak we can alternatively use (as they did as well)
the first-order perturbation approximation which leads to:
$\phi_s(x,X,t)\simeq \phi_n(x)=\sqrt{(\frac{2}{L})}\sin{(n\pi x/L)}$
and $E_{n,P}(X,t)-\frac{P^2}{2M}=(\hbar n\pi/L)^2/(2m)+\delta E$
with
\begin{eqnarray}
\delta E=-\hbar\epsilon g(t)X\langle
n|\delta(\hat{x}-x_0)|n\rangle=-\hbar\epsilon g(t)X
|\phi_n(x_0)|^2.\label{15}
\end{eqnarray}
We point out that there is actually a small slope discontinuity at
$x_0$ since for the 1D Green function we must have:
\begin{eqnarray}
d\phi_s(x,X,t)/dx|_{x_0+\delta}-d\phi_s(x,X,t)/dx|_{x_0-\delta}\nonumber\\=-2m\epsilon
g(t)X\phi_s(x_0,X,t)/\hbar\end{eqnarray} with $\delta\rightarrow
0^{+}$. In the weak coupling regime we can neglect this effect and
therefore the cavity mode can fairly be considered as `protected'.
The next step is to expand the full system wave function by solving
the Schr\"{o}dinger equation $i\hbar d\Psi(t) dt=H(t)\Psi(t)$ and
using these eigenmodes labeled by the index $n$ of the cavity mode
(here we will limit our analysis to $n=1$) and $P$ the `orthodox'
momentum of the pointer. We have :
\begin{eqnarray}\Psi(x,X,t)=\Sigma_{n,P}b_{n,P}(t,X)\Psi_{n,P}(x,X,t).\end{eqnarray} In
the adiabatic approximation we write the amplitude coefficients as
\begin{eqnarray}b_{n,P}(t,X)=c_{n,P}(t,X)e^{-i\int^t_{-\infty}dt'E_{n,P}(X,t')/\hbar}\end{eqnarray}
and we get here:
\begin{eqnarray}
\Psi(x,X,t)\simeq\phi_s(x,X,t)\int
\frac{dP}{\sqrt{(2\pi\hbar)}}M(P)e^{iPX/\hbar}e^{-i\frac{P^2}{2M\hbar}t}\nonumber\\
\cdot e^{i\int^t_{-\infty}dt' [\epsilon g(t')X
|\phi_n(x_0)|^2-\beta_n(t')]}e^{-iE_nt/\hbar}
\end{eqnarray} with $i\beta_n(t)=\langle
n|\frac{d}{dt}|n\rangle\simeq 0$. In doing this calculation we
supposed the initial state begin unentangled, i.e., like for the von
Neumann procedure. This initial state corresponds to the product of
a undisturbed cavity mode $n=1$ (i.e. $\phi_s(x,X,t)\simeq
\sqrt{(\frac{2}{L})}\sin{(\pi x/L)}$) by a localized wave packet
with gaussian Fourier coefficient $M(P)\propto e^{-\Delta p^2}$. The
coupling is supposed to be slow and weak so that the energy given by
the interaction is not large enough to induce transition between
different eigenmodes. For $t\rightarrow+\infty$ we thus get
\begin{equation}
\Psi(x,X,t)\simeq\phi_s(x,X,t)e^{-iE_nt/\hbar}\cdot\Psi_M(X,t)\cdot
e^{i\epsilon X |\phi_n(x_0)|^2}\label{20}
\end{equation} which shows that the main result of the interaction
is to induce a phase kick to the pointer wave packet $\Psi_M(X,t)$.
If we neglect the free space spreading of the pointer wave packet
this phase shift will impose a translation $\Delta
P\simeq+\hbar\epsilon|\phi_n(x_0)|^2$ in the Fourier space such as
$M'(P)=M(P-\Delta P)$. This results in a protective measurement
where the local adiabatic coupling keeps the confined
mode $\phi_s(x,X,t)$ undisturbed.\\
\indent Now comes the paradox: since the cavity mode is protected
and since it corresponds to a de Broglie vanishing velocity of the
particle S (i.e., $dx(t)/dt=\partial_x S(\mathbf{x},t)/m=0$) we
deduce that the pointer M is disturbed by the local interaction
centered at $x=x_0$ even though S never approaches this position.
How could that be? For Aharonov et al. one can hardly avoid the
conclusion that Bohmian trajectories are just a mathematical
construct. The same conclusion was actually given (although in a
less technical way) in a previous paper \cite{Aharonov1996} were the
authors concluded that Bohmian trajectory contradicts the natural
statement: `an empty wave should not yield observable effects on
other particles'. Indeed, the measuring device recording
$|\psi_n(x_0)|^2$ in the `empty' region surrounding $x_0$ yields
non-null outcomes (identical conclusions were discussed in
\cite{Erez2004}). In his review paper \cite{Vaidman2005} Vaidman
however considered the problem from a wider perspective and
commented that for him in the framework of Bohmian mechanics there
is no fundamental problem since `these experiments are \emph{not}
good verification measurements' so that Bohmian proponents have `a
good defense'. Nevertheless, this looks mysterious or magical since
one would like to find where does the force acting on the pointer
come from? Furthermore, even if one is not accepting the ontology
proposed by de Broglie's pilot-wave it was at least possible until
now to accept its self-consistency. Does protective measurement
changes the rules? Indeed, magical forces have no place in physics.
In order to remove some of these ambiguities and magical features I
developed in a paper published in 2005 \cite{Drezet2005} a dynamical
analysis of the protective measurement discussed
in~\cite{Aharonov1999} seen from the point of view of pilot-wave
theory. I will now summarize my reasoning using the calculations
given before. First, we observe that the quantum potential for the
system given by Eq.~\ref{20} is :
\begin{eqnarray}
Q(x,X,t)=\frac{-\hbar^2}{2m}\frac{\frac{\partial^2}{\partial
x^2}|\Psi(x,X,t)|}{|\Psi(x,X,t)|}+\frac{-\hbar^2}{2M}\frac{\frac{\partial^2}{\partial
X^2}|\Psi(x,X,t)|}{|\Psi(x,X,t)|} \nonumber\\
\simeq\frac{-\hbar^2}{2m}\frac{\frac{\partial^2}{\partial
x^2}|\phi_s(x,X,t)|}{|\phi_s(x,X,t)|}+\frac{-\hbar^2}{2M}\frac{\frac{\partial^2}{\partial
X^2}|\Psi_M(X,t)|}{|\Psi_M(X,t)|}.
\end{eqnarray} Here, we fairly neglected the small contributions of terms containing the $X$
derivatives of $|\phi_s(x,X,t)|$. Now using Eqs.~\ref{14}, \ref{15}
and the fact that $\phi_s(x,X,t)\simeq \phi_1(x)$ is real  we
immediately get
\begin{eqnarray}
Q(x,X,t)\simeq\frac{(\hbar\pi/L)^2}{2m}-\hbar\epsilon g(t)X
|\phi_1(x_0)|^2\nonumber\\+\hbar\epsilon
g(t)\delta(x-x_0)X+\frac{-\hbar^2}{2M}\frac{\frac{\partial^2}{\partial
X^2}|\Psi_M(X,t)|}{|\Psi_M(X,t)|}.
\end{eqnarray} Now, the potential acting in the Hamilton-Jacobi
equation is $U=V+Q$ where $V=-\hbar\epsilon g(t)\delta(x-x_0)X$ is
the `classical' local interaction potential associated with the
protective measurement protocol. Here, this leads therefore to
\begin{eqnarray}
U(x,X,t)\simeq\frac{(\hbar\pi/L)^2}{2m}-\hbar\epsilon g(t)X
|\phi_1(x_0)|^2+\frac{-\hbar^2}{2M}\frac{\frac{\partial^2}{\partial
X^2}|\Psi_M(X,t)|}{|\Psi_M(X,t)|}.
\end{eqnarray} Remarkably, the local potential has been removed
from the total Hamiltonian because the singular term in $V$ exactly
compensates the one in $Q$. This implies that from the framework of
pilot-wave theory the interaction is highly quantum-like, i.e., it
has not classical analog. This is even more clear   in the Newton
picture.  Newton's law reads indeed $md^2x(t)/dt^2=F_x$ and
$md^2X(t)/dt^2=F_X$ and with the definition for $U$ this implies for
the evolution of $S$:
\begin{eqnarray}
F_x=-\frac{\partial}{\partial x }U(x,X,t)\simeq0\end{eqnarray} i.e.
the force applied on the Bohmian particle vanishes.  This situation
is exactly similar to the one obtained in the Einstein example or in
s state atom discussed by de Broglie, Pauli and Einstein. In each
cases the quantum potential is constant over the region of interest
so that the particle can indeed stay in static equilibrium in full
agreement with the de Broglie guidance condition
$mdx(t)/dt=\partial_x S(x,X,t)=0$. The big difference is that in the
protective measurement there is actually a local force
$-\frac{\partial}{\partial x }V$ but its effect is compensated by an
additional quantum term in $-\frac{\partial}{\partial x }Q$.
Remarkably, the situation is completely different for  the meter $M$
since we get:
\begin{eqnarray}
F_X\simeq-\frac{\partial}{\partial X }U(x,X,t)\simeq+\hbar\epsilon
g(t) |\phi_1(x_0)|^2
\end{eqnarray} in agreement with the momentum kick $\Delta P =\int dt' F_X(t')=+\hbar\epsilon
 |\phi_1(x_0)|^2$ introduced previously. Therefore, the pointer
deviation is completely justified from the point of view of de
Broglie and Bohm approach. However, here the force applied on $M$ is
of quantum origin and not the local and classical term
$-\frac{\partial}{\partial x }V$.
\section{A short conclusion}
Finally, what can we deduce from this story? We reviewed pilot-wave
theory and showed that the surrealism objection is very old and goes
back to the origin of the theory. Einstein did not like this theory
in part because the trajectories predicted in general don't follow
our classical intuitions about dynamics. Latter, this surrealism was
criticized because very often even causality is affected by
pilot-wave. This of course included non locality as studied by Bell
but also modifications of our intuitions about what should a
trajectory in an interferometer be. The work by Aharonov et al. on
protective measurements follows this strategy, and indeed, it
confirms that pilot-wave is not classical. Still, this theory is the
only known quantum ontology (Lev Vaidman will certainly not agree
here) which is completely self consistent at the mathematical level
and at the same time explains every experimental fact (too many
words could be said here about the Everett's
interpretation~\cite{Everett} and its problems associated with
probabilities and this will be therefore omitted). Of course, it is
probably only a temporary expedient and pilot-wave theory has no
convincing or univocal relativistic generalization, but  to quote
Bell `Should it not be taught, not as the only way, but as an
antidote to the prevailing
complacency?'~\cite{Bell} (see p. 160).\\

The author thanks  Serge Huant for helpful suggestions during the
preparation of the manuscript.

\end{document}